\documentclass{sig-alternate}

\newcommand{\ourtitle}{Controversy and Sentiment in Online News}

\usepackage[hide]{chato-notes}
\pdfoutput=1

\permission{\inote{This version generated on: \today}}
\copyrightetc{Computation+Journalism Symposium 2014.\\October 24-25, 2014. Columbia University, New York, NY, USA}

\def\pprw{8.5in} 
\def\pprh{11in}

\usepackage{balance}       
\usepackage{graphics}      
\usepackage{times}         
\usepackage[hyphens]{url}  
\usepackage{paralist}      
\usepackage{booktabs}      
\usepackage[hang]{subfigure}     

\usepackage[square, comma, sort&compress, numbers]{natbib} 
\usepackage{csquotes}\MakeOuterQuote{"}                    

\newcommand{\spara}[1]{\smallskip\noindent{\bf #1}}

\newcommand{\term}[1]{{\em #1}}

\title{\ourtitle}
\numberofauthors{1}
\author{
Yelena Mejova$^{*}$,
Amy X. Zhang$^{\ddag}$,
Nicholas Diakopoulos$^{\dag}$,
Carlos Castillo$^{*}$\\
\affaddr{
$^{*}$Qatar Computing Research Institute,
$^{\dag}$University of Maryland,
$^{\ddag}$MIT CSAIL}\\
\email{
ymejova@qf.org.qa,
axz@mit.edu,
nicholas.diakopoulos@gmail.com,
chato@acm.org}
}

\begin{document}
\maketitle


\begin{abstract}
How do news sources tackle controversial issues? In this work, we take a data-driven approach to understand how controversy interplays with emotional expression and biased language in the news. We begin by introducing a new dataset of controversial and non-controversial terms collected using crowdsourcing.
Then, focusing on 15 major U.S. news outlets, we compare millions of articles discussing controversial and non-controversial issues over a span of 7 months.
We find that in general, when it comes to controversial issues, the use of negative affect and biased language is prevalent, while the use of strong emotion is tempered. We also observe many differences across news sources.
Using these findings, we show that we can indicate to what extent an issue is controversial, by comparing it with other issues in terms of how they are portrayed across different media.
\end{abstract}

\section{Introduction}\label{sec:introduction}

One of the most vital functions of the news media is to serve as a place to critically examine and present information about social, political, economic, and ideological issues of the day. Many of these issues are controversial, in the sense that they provoke arguments in which people express strong opposing opinions~\cite{pennacchiotti_2010_controversies}.

For this reason, journalists must often take special precaution and make careful language choices when they write about controversial issues.
This can often manifest as different ways of using language, for instance news sources will often use a series of terms to signal that a controversy exists, such as ``outcry,'' ``furor,'' and  ``uproar''~\cite{cramer_2011_controversy_event_category}. 
%
It has been theorized that journalists can also become susceptible to the ideologies, attitudes, and pressures of their organization~\cite{scheufele1999framing}, as well as unstated rules and norms~\cite{entman2007framing}. These inputs can influence the particular language used to discuss controversial issues within a particular news source. The difference in framing could be subtle enough to be unnoticeable to the casual reader. However, using computational techniques in textual analysis, we can analyze large datasets of articles for consistent differences in the way different news organizations write about controversial issues.

\spara{Our contribution.} In this work, we quantify the use of emotional and biased language when presenting controversial issues in the news. We begin by building a list of controversial and non-controversial terms in current news in the U.S. using crowdsourcing techniques. Then, we perform a large-scale analysis of millions of news articles from 15 U.S.-based news sources. We focus on the expression of sentiment using a series of lexical resources containing words conveying positive and negative emotions; this automatic analysis helps reduce the inherent subjectivity of traditional content analysis methods.

We demonstrate that controversial issues in news can be characterized by the use of fewer positive words and a greater presence of negative words. This finding is consistent across different media sources and confirmed with 4 different sentiment lexicons. Interestingly, we find that the use of highly emotional terms (as opposed to mild ones) is \emph{less} likely in the context of controversial topics, suggesting a self-moderation on the part of the news sources.

Additionally, we perform an analysis based on a vocabulary of words signaling bias obtained from discussions in Wikipedia, and find that these bias terms also tend to occur more frequently in articles mentioning controversial topics, and can serve as a fairly accurate predictor of the level of controversy.

\medskip
The next section outlines previous work related to controversy in news media.
Next, we describe our dataset of online news (Section~\ref{sec:news_dataset}) and
describe the process we used to label strongly controversial, somewhat controversial, and non-controversial words (Section~\ref{sec:labeling}).
We then compare controversial and non-controversial articles in terms of a series of bias and sentiment lexicons in Section~\ref{sec:sentiments-controversy},
and discuss the differences in the strength with which annotators perceive a topic as controversial and the treatment it received in news media in Section~\ref{sec:ranking-words}.
Lastly we discuss the implications and limitations of such computational approaches to media analysis.

\section{Previous Work}

Controversy has been examined in both social media, including Wikipedia and Twitter, and more traditional news sites.

The unique structure of Wikipedia as a collaborative endeavor has been used by \citet{rad_2012_controversial} who detect controversy based on mutual reverts, bi-polarity in the collaboration network, and/or mutually-reinforced scores for editors and articles (controversial editors work in controversial articles).
The fact that an article on Wikipedia is controversial has then been used to evaluate the level of controversy of other documents (e.g. web pages) by mapping them to related Wikipedia articles~\cite{hacohen_2013_controversy}. 

Taking a content-driven approach, \citet{pennacchiotti_2010_controversies} detect controversies around celebrities in Twitter. 
They use a number of features including the presence of sentiment-bearing words, swear words, and words in a list of controversial topics from Wikipedia.

\spara{Controversies in online news.}
In news sites, \citet{choi_2010_controversial} detect controversial issues by looking at words that frequently appear in contexts containing positive and/or negative sentiment words. 
Also using lexicons, \citet{chimmalgi_2010_controversy} study controversy in user comments of news articles. 
Taking a more data-driven approach,~\citet{awadallah_2012_harmony} describe a method in which opinion holders and their opinions in news articles are identified by an iterative method based on a seed set of patterns that describe expressions of support or opposition to an idea. 

\medskip
In contrast to previous work that applies sentiment analysis to controversies, we do not assume that sentiments and controversy are related, but instead demonstrate it experimentally, thus not only providing an empirical basis for future use of sentiment lexicons, but also discovering new insight into potential self-moderation of the news sources. 
%


\section{News Articles}\label{sec:news_dataset}

Data was provided by NewsCred,\footnote{\url{http://newscred.com/}} which aggregates news content from thousands of news sources, and makes it available via an API (described further in \citet{diakopoulos_2013_visual}). 
We chose 15 high-volume sources from NewsCred and considered their articles in the period from March to September 2013. The main criterion we used for this selection was variety---including both national and regional sources---while at the same time we attempted to keep the list relatively brief for ease of analysis and exposition:
\begin{inparaenum}[]
\item CNN,
\item Reuters,
\item Usa Today,
\item Los Angeles Times,
\item Washington Post,
\item Chicago Tribune,
\item News Day,
\item Minneapolis Star Tribune,
\item Houston Chronicle,
\item Philadelphia Inquirer,
\item Honolulu Star-Advertiser,
\item Huffington Post,
\item New York Times,
\item Pro\-Publica, and
\item Talk\-ing Points Memo.
\end{inparaenum}

Two data gathering tasks were done.
First, we collected a random subset of articles for the purposes of obtaining high-frequency words. This is the list of initial topics used for the labeling task described in Section~\ref{sec:labeling}.
Second, we searched the selected media sources for articles containing each topic word using the NewsCred API. In total, over 21 million articles were collected, with an average of around 3,000 and median of 1,000 per topic per news source. Reuters was the most prolific source with an average of 11,659 articles per topic, and ProPublica the least at 31 articles.


\section{Controversial Words}\label{sec:labeling}

Whether a word is controversial or not is a highly subjective and context-dependent matter. Since controversy is socially constructed, we performed an annotation effort with a relatively large pool of annotators. We employed 25 annotators based in the U.S., hired through crowdsourcing platform CrowdFlower (\url{https://crowd-flower.com/}) and paid \$0.80 (USD) for every 100 words annotated, following standard pricing practices of this platform.\footnote{In total 40 annotators participated in the task, but only 25 of them contributed more than 100 labels. This is common in crowdsourcing platforms.}

\spara{Initial Words.} 
The initial list was composed of the top frequent words in a large random sample of articles. We filtered this list to keep only single-word nouns using a part-of-speech tagger.
Next, we removed generic English stopwords, as well as frequent news-specific stopwords (\term{ap}, \term{broadcast}, \term{press}, \term{published}, \term{rewritten}, \term{redistributed}, \term{rights}, \term{copyright}, \term{reserved}), and kept the top 2,000 most frequent terms.

\spara{Crowdsourced annotation.}
We used a four-point scale to classify 2,000 high-frequency terms, asking 7 ``trusted'' annotators per word the following question:
\begin{quote}\small
\mbox{You need to be familiar with U.S. news media. Indicate if {\em word} is:}
\begin{compactitem}
\item[(C3)] Strongly Controversial: people often disagree and debate with opposing viewpoints.
\item[(C2)] Somewhat Controversial: people sometimes disagree or have debates with opposing viewpoints.
\item[(C1)] Less Controversial: people infrequently disagree or have debates with opposing viewpoints.
\item[(C0)] Non-Controversial: people almost never disagree or debate with opposing viewpoints.
\end{compactitem}
\end{quote}
``Trusted'' annotators are selected by including a set of terms for which the label was known. These terms were obtained by a preliminary task in which we asked 5 crowdsourcing workers for each of 1,000 words whether they believed it to be controversial in U.S. news media or not. We selected a balanced set of 94 words for which there was perfect agreement in the preliminary task and assigned them to C0 and C1 (for non-controversial terms) and C2 and C3 (for controversial terms). Annotators who did not agree substantially with this gold standard were not considered ``trusted.''


\begin{table}
\caption{List of words identified during the crowdsourcing task.}
\label{tbl:words}
\centering\scriptsize\begin{tabular}{p{\columnwidth}}
\toprule
{\bf Strongly Controversial (145):}
\term{abuse}, \term{administration}, \term{afghanistan}, \term{aid}, \term{america}, \term{american}, \term{army}, \term{attack}, \term{attacks}, \term{authorities}, \term{authority}, \term{ban}, \term{banks}, \term{benefits}, \term{bill}, \term{bills}, \term{border}, \term{budget}, \term{campaign}, \term{candidate}, \term{candidates}, \term{catholic}, \term{china}, \term{chinese}, \term{church},
\term{concerns}, \term{congress}, \term{conservative}, \term{control}, \term{country}, \term{court}, \term{crime}, \term{criminal}, \term{crisis}, \term{cuts}, \term{debate}, \term{debt}, \term{defense}, \term{deficit}, \term{democrats}, \term{disease}, \term{dollar}, \term{drug}, \term{drugs}, \term{economy}, \term{education}, \term{egypt}, \term{election}, \term{elections}, \term{enforcement},
\term{fighting}, \term{finance}, \term{fiscal}, \term{force}, \term{funding}, \term{gas}, \term{government}, \term{gun}, \term{health}, \term{immigration}, \term{inaccuracies}, \term{india}, \term{insurance}, \term{investigation}, \term{investigators}, \term{iran}, \term{israel}, \term{job}, \term{jobs}, \term{judge}, \term{justice}, \term{killing}, \term{korea}, \term{labor}, \term{land}, \term{law},
\term{lawmakers}, \term{laws}, \term{lawsuit}, \term{leadership}, \term{legislation}, \term{marriage}, \term{media}, \term{mexico}, \term{military}, \term{money}, \term{murder}, \term{nation}, \term{nations}, \term{news}, \term{obama}, \term{offensive}, \term{officials}, \term{oil}, \term{parties}, \term{peace}, \term{police}, \term{policies}, \term{policy}, \term{politics}, \term{poll}, \term{power},
\term{president}, \term{prices}, \term{primary}, \term{prison}, \term{progress}, \term{race}, \term{reform}, \term{republican}, \term{republicans}, \term{restrictions}, \term{rule}, \term{rules}, \term{ruling}, \term{russia}, \term{russian}, \term{school}, \term{security}, \term{senate}, \term{sex}, \term{shooting}, \term{society}, \term{spending}, \term{strategy}, \term{strike}, \term{support}, \term{syria}, \term{syrian}, \term{tax}, \term{taxes}, \term{threat}, \term{trial}, \term{unemployment}, \term{union}, \term{usa}, \term{victim}, \term{victims}, \term{violence}, \term{vote}, \term{voters}, \term{war}, \term{washington}, \term{weapons}, \term{world} \\
\midrule
{\bf Somewhat Controversial (45):}
\term{account}, \term{advantage}, \term{amount}, \term{attorney}, \term{chairman}, \term{charge}, \term{charges}, \term{cities}, \term{class}, \term{comment}, \term{companies}, \term{cost}, \term{credit}, \term{delays}, \term{effect}, \term{expectations}, \term{families}, \term{family}, \term{february}, \term{germany}, \term{goal}, \term{housing}, \term{information}, \term{investment}, \term{markets},
\term{numbers}, \term{oklahoma}, \term{parents}, \term{patients}, \term{population}, \term{price}, \term{projects}, \term{raise}, \term{rate}, \term{reason}, \term{sales}, \term{schools}, \term{sector}, \term{shot}, \term{source}, \term{sources}, \term{status}, \term{stock}, \term{store}, \term{worth}\\
\midrule
{\bf Non-Controversial (272): }
\term{60s}, \term{70s}, \term{addition}, \term{address}, \term{afternoon}, \term{agreed}, \term{amp}, \term{angeles}, \term{answer}, \term{april}, \term{attention}, \term{avenue}, \term{average}, \term{ball}, \term{base}, \term{bay}, \term{beach}, \term{beginning}, \term{bit}, \term{block}, \term{blue}, \term{bowl}, \term{box}, \term{boy}, \term{boys}, \term{brother}, \term{building}, \term{bus},
\term{call}, \term{calling}, \term{calls}, \term{camp}, \term{car}, \term{cars}, \term{central}, \term{cents}, \term{click}, \term{close}, \term{cloudy}, \term{club}, \term{coast}, \term{cup}, \term{dallas}, \term{date}, \term{daughter}, \term{davis}, \term{day}, \term{decade}, \term{decades}, \term{december}, \term{def}, \term{delivery}, \term{door}, \term{download}, \term{drive}, \term{eagles},
\term{end}, \term{entire}, \term{era}, \term{evening}, \term{face}, \term{faces}, \term{facility}, \term{fall}, \term{fans}, \term{father}, \term{feel}, \term{feeling}, \term{feet}, \term{fell}, \term{field}, \term{finish}, \term{floor}, \term{form}, \term{fort}, \term{francisco}, \term{friday}, \term{friend}, \term{friends}, \term{fun}, \term{girl}, \term{girls}, \term{ground}, \term{gt}, \term{guy},
\term{guys}, \term{half}, \term{hall}, \term{hand}, \term{hands}, \term{hawaii}, \term{heart}, \term{heat}, \term{heavy}, \term{hill}, \term{hits}, \term{hold}, \term{hopes}, \term{host}, \term{hotel}, \term{hour}, \term{hours}, \term{house}, \term{houston}, \term{hundreds}, \term{husband}, \term{ice}, \term{illinois}, \term{index}, \term{indiana}, \term{innings}, \term{island}, \term{january},
\term{johnson}, \term{jones}, \term{june}, \term{kansas}, \term{kind}, \term{lack}, \term{lake}, \term{leave}, \term{lee}, \term{letter}, \term{levels}, \term{light}, \term{line}, \term{lines}, \term{lot}, \term{lows}, \term{lt}, \term{main}, \term{make}, \term{makes}, \term{mark}, \term{mass}, \term{material}, \term{matter}, \term{medium}, \term{men}, \term{mid}, \term{middle}, \term{miles},
\term{mind}, \term{minneapolis}, \term{minutes}, \term{moment}, \term{monday}, \term{month}, \term{months}, \term{morning}, \term{mother}, \term{mountain}, \term{move}, \term{mph}, \term{museum}, \term{names}, \term{natural}, \term{net}, \term{night}, \term{north}, \term{note}, \term{notes}, \term{november}, \term{october}, \term{opening}, \term{park}, \term{part}, \term{parts}, \term{pass},
\term{period}, \term{person}, \term{philadelphia}, \term{pick}, \term{pitch}, \term{plant}, \term{play}, \term{player}, \term{playing}, \term{pm}, \term{point}, \term{post}, \term{practice}, \term{put}, \term{quarter}, \term{rain}, \term{read}, \term{reading}, \term{red}, \term{rest}, \term{restaurant}, \term{rise}, \term{rock}, \term{rose}, \term{round}, \term{sale}, \term{san}, \term{saturday},
\term{scene}, \term{search}, \term{season}, \term{seasons}, \term{seconds}, \term{selling}, \term{september}, \term{series}, \term{set}, \term{showers}, \term{showing}, \term{shows}, \term{sign}, \term{signs}, \term{smith}, \term{son}, \term{sox}, \term{special}, \term{spot}, \term{spring}, \term{square}, \term{stadium}, \term{stage}, \term{start}, \term{starting}, \term{starts}, \term{station},
\term{stay}, \term{step}, \term{stores}, \term{street}, \term{student}, \term{summer}, \term{sun}, \term{sunday}, \term{thing}, \term{things}, \term{thinking}, \term{thought}, \term{thousands}, \term{thunderstorms}, \term{thursday}, \term{time}, \term{title}, \term{top}, \term{total}, \term{transportation}, \term{type}, \term{unit}, \term{valley}, \term{vehicle}, \term{version}, \term{village},
\term{visit}, \term{wait}, \term{walk}, \term{wall}, \term{watch}, \term{water}, \term{ways}, \term{wednesday}, \term{week}, \term{weekend}, \term{weeks}, \term{williams}, \term{wind}, \term{winds}, \term{winner}, \term{winter}, \term{word}, \term{writer}, \term{yards}, \term{year}, \term{years}, \term{york} \\
\bottomrule
\end{tabular}
\end{table}

Finally, we considered only labels for which the majority label was larger than 60\% among the 7 workers. This yields Table~\ref{tbl:words} containing 145 controversial terms having label C3, 45 terms having label C2, and 272 terms having label C0. Class C1 is made of borderline cases and elicited very little agreement, it was thus not considered.

Controversial terms in Table~\ref{tbl:words} are part political terms referring to \term{congress} and \term{legislation}, social terms like \term{education} and \term{unemployment}, country names including \term{russia} and \term{china}, and terms such as \term{criminal} and \term{threat}. Those in the medium category may potentially belong to a controversial topic, such as \term{february} being Black History Month in the U.S. Finally, the non-controversial terms include mostly generic words.


\section{Controversy and Sentiment}\label{sec:sentiments-controversy}

Using the dataset we describe above, we explore the vocabulary used in articles mentioning controversial/non-controversial topics using a series of lexical resources related to sentiment and bias.

\spara{Method.}
The general methodology is to consider in turn each word $w$ in our vocabulary of 462 topics, and each media source $s$ in our list of 15 news media. Then, collect all articles in the source $s$ containing the word $w$, eliminate duplicate and near-duplicate articles (typically a by-product of articles that have multiple URLs), and combine all the articles in one topical (and very large) super-article for analysis.

The analysis consists of examining this content using a lexical resource, counting the proportion of words matching a certain category in the lexical resource.
We consider the following sentiment lexicons:

\begin{compactenum}
\item[a.] {Affective Norms for English Words}\footnote{\url{http://csea.phhp.ufl.edu/media/anewmessage.html}} (ANEW) is a set of normative emotional ratings for 2,476 English words. We use the ``valence'' rating considering positive (respectively, negative) the ratings above (respectively, below) the mean.
\item[b.] {General Inquirer}\footnote{\url{http://www.wjh.harvard.edu/~inquirer/homecat.htm}} is a list of 1,915 words classified as positive, and 2,291 words classified as negative.
\item[c.] {MicroWNOp}\footnote{\url{http://www-3.unipv.it/wnop/}}~\cite{cerini2007micrownop} is a list of 1,105 WordNet \emph{synsets} (cognitive synonyms) classified as positive, negative, or neutral.
\item[d.] {SentiWordNet}\footnote{\url{http://sentiwordnet.isti.cnr.it/}}~\cite{baccianella2010sentiwordnet} assigns to each synset of WordNet (around 117,000) a positive and negative score determined by a diffusion process.
\end{compactenum}

\noindent Additionally, we use a bias-specific lexicon:

\begin{compactitem}
 \item[e.] {Bias Lexicon}\footnote{\url{http://www.mpi-sws.org/~cristian/Biased_language.html}} is a list of 654 bias-related lemmas extracted from the edit history of Wikipedia by \citet{recasens2013linguistic}. Sentiment words are used as contributing features in the construction of this bias lexicon.
\end{compactitem}

To assess how prominent bias- and sentiment-laden terms are in controversial topics, we count the number of times each lexicon term is used in each super-article, and divide this count by the total length of the article, resulting in a proportion of the text which uses the lexicon terms. 

\begin{figure}[b]
\vspace{-0.3cm}
\centering\includegraphics[width=0.36\textwidth,natwidth=468,natheight=1345]{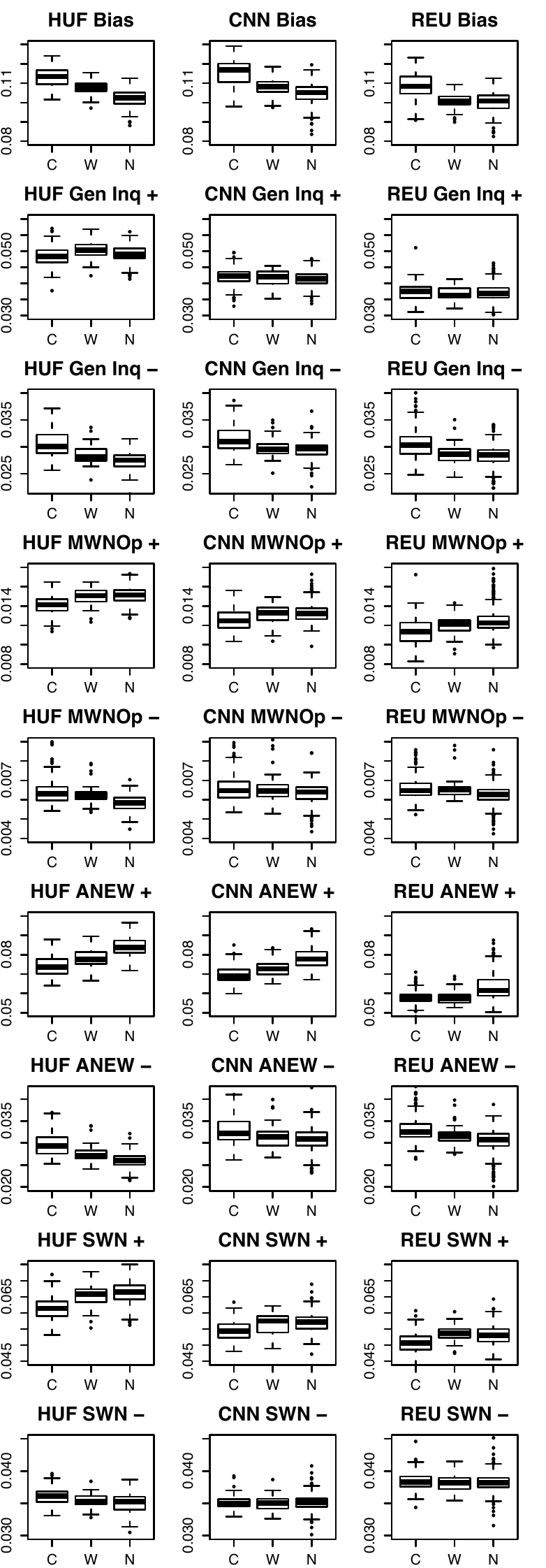}
\caption{Distributions of the proportion of biased, positive ($+$), and negative ($-$) words in the Huffington Post (HUF), CNN, and Reuters (REU), across controversial topics (C), somewhat controversial topics (W) and non-controversial topics (N).}
\label{fig:boxplots}
\end{figure}

\spara{Bias, positive and negative emotions.}
The distributions of the proportion of words matching each vocabulary are shown in Figure~\ref{fig:boxplots} in which for brevity we have selected three media: Huffington Post (HUF), CNN, and Reuters (REU). As we show in the next section, these three news media sources differ in the extent to which the use of lexicon words are used around controversial words in their articles (from most to least).

First, we observe that {\em the use of bias terms is more likely in controversial topics than non-controversial topics}. This is statistically significant at $p<0.01$ for all 15 news sources.

Second, {\em the use of negative terms is more likely in controversial topics than non-controversial topics}. This is statistically significant at $p<0.01$ for 46 out of the 60 combinations of source and lexicon (15 sources and 4 sentiment lexicons), with 9 ties (no difference significant at this level in either direction) and 5 cases in which the difference is significant at this level in the opposite direction.

Third, {\em the use of positive terms is more likely in non-controversial topics than controversial topics}. This is statistically significant at $p<0.01$ for 40 out of the 60 conditions, with 16 ties and 4 cases with a significant difference in the opposite direction.

\spara{Strong emotions.}
Three of our lexicons (ANEW, MicroWNOp and SentiWordNet) include scores that allow us to distinguish between weakly and strongly emotional terms.
We observe that {\em the use of strong emotional words is less likely in controversial topics}. This is statistically significant at $p<0.01$ for 32 out of 45 conditions (15 sources and 3 sentiment lexicons), with only 1 of the remaining conditions having a significant difference in the other direction.

\spara{Differences across sources.}
We find a great variety in the different treatment that controversial and non-controversial topics have, in terms of the use of biased and emotional words.
In terms of statistical significance, the clearer difference between controversial and non-controversial topics was observed using
\begin{inparaenum}[(i)]
\item the lexicon of bias words, 
\item the General Inquirer strong sentiment words, and 
\item the ANEW negative words.
\end{inparaenum}

We next rank the media sources in terms of their different usage of words in these lexicons in controversial and non-controversial topics. The top 5 sources for each one are shown on Table~\ref{tbl:agencyranking}. We note that several media sources repeat in this list, with Huffington Post, Washington Post, and New York Times remaining on top, indicating a consistently large difference in their usage of sentiment words around controversial topics, compared to non-controversial ones.

\begin{table}
\caption{Top news sources in terms of the difference in the use of bias/emotional words in controversial and non-controversial topics.}
\label{tbl:agencyranking}
\begin{center}
{\footnotesize
\begin{tabular}{lccc}\toprule
& Bias & Gen Inq $Strong$ & ANEW $-$ \\ \midrule
1. & Huffington Post         &       Huffington Post         &       Huffington Post \\
2. & Washington Post         &       Washington Post         &       USA Today \\
3. & New York Times  &       New York Times  &       New York Times \\
4. & LA Times        &       CNN     &       Washington Post \\
5. & USA Today       &       LA Times        &       LA Times \\
\bottomrule
\end{tabular}}
\end{center}
\end{table}

\section{Ranking Controversy Words}\label{sec:ranking-words}

Finally, we assign a score to each topic in our list of controversial and non-controversial terms by using logistic regression, using as input features the proportion of words from each lexicon in each news source, and as training data the manually-labeled words (using only the classes C3 and C0 that represent the extreme values). This is done using feature selection, selecting 5 features out of the total 195, and training a logistic regression classifier.
Then, the same classifier is applied to the training data (the purpose of this is not to generalize to unseen topics, but to understand the existing one), and a score between 0 and 1 is computed for each word (0.0 is non-controversial, 0.5 is undecided, and 1.0 is strongly controversial).

\begin{figure}
\centering
\subfigure[Non-controversial words; correctly classified words appear below the horizontal line.\label{fig:word-scores-nc}]{\includegraphics[width=\columnwidth]{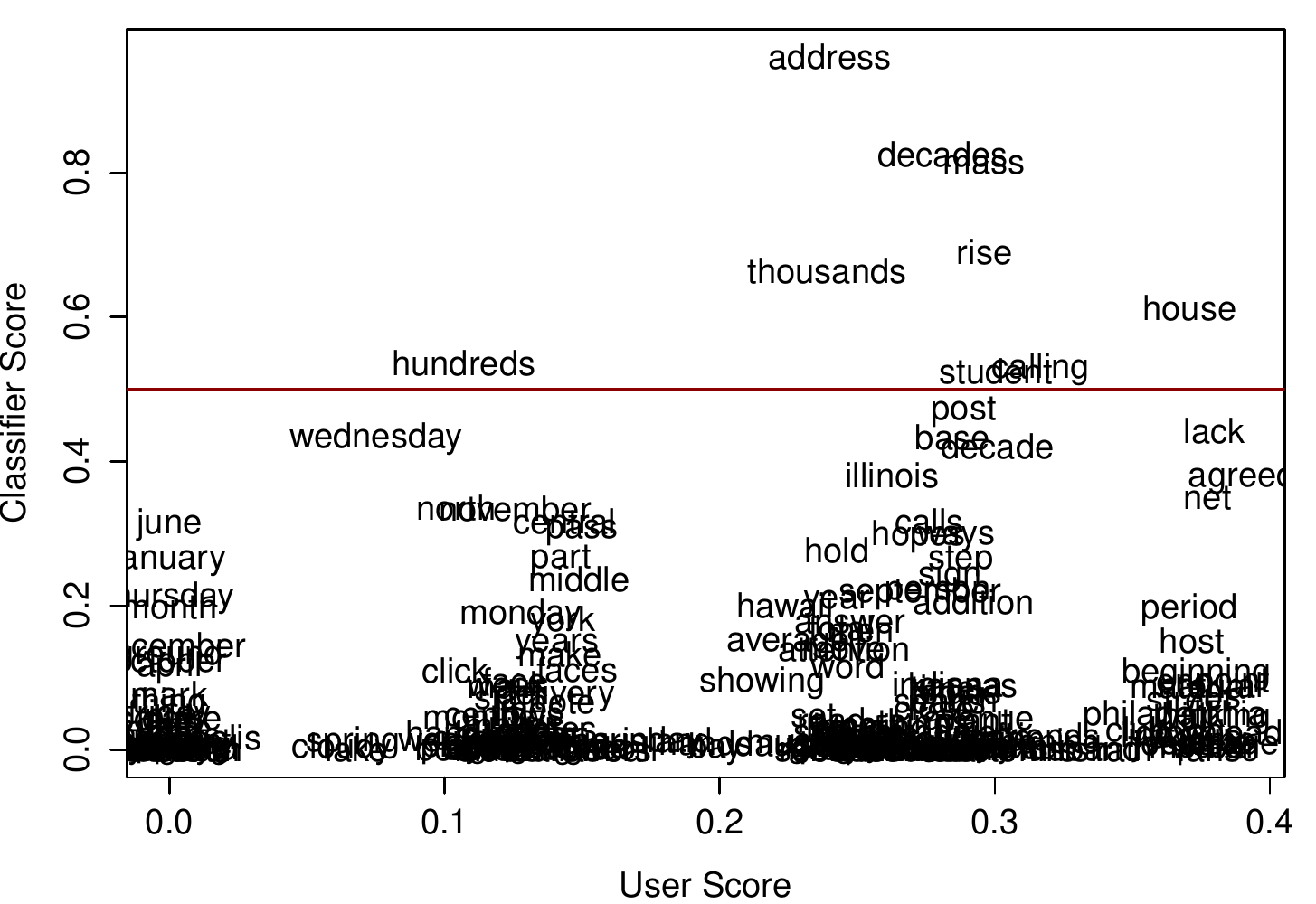}}
\subfigure[Controversial words; correctly classified words appear above the horizontal line.\label{fig:word-scores-c}]{\includegraphics[width=\columnwidth]{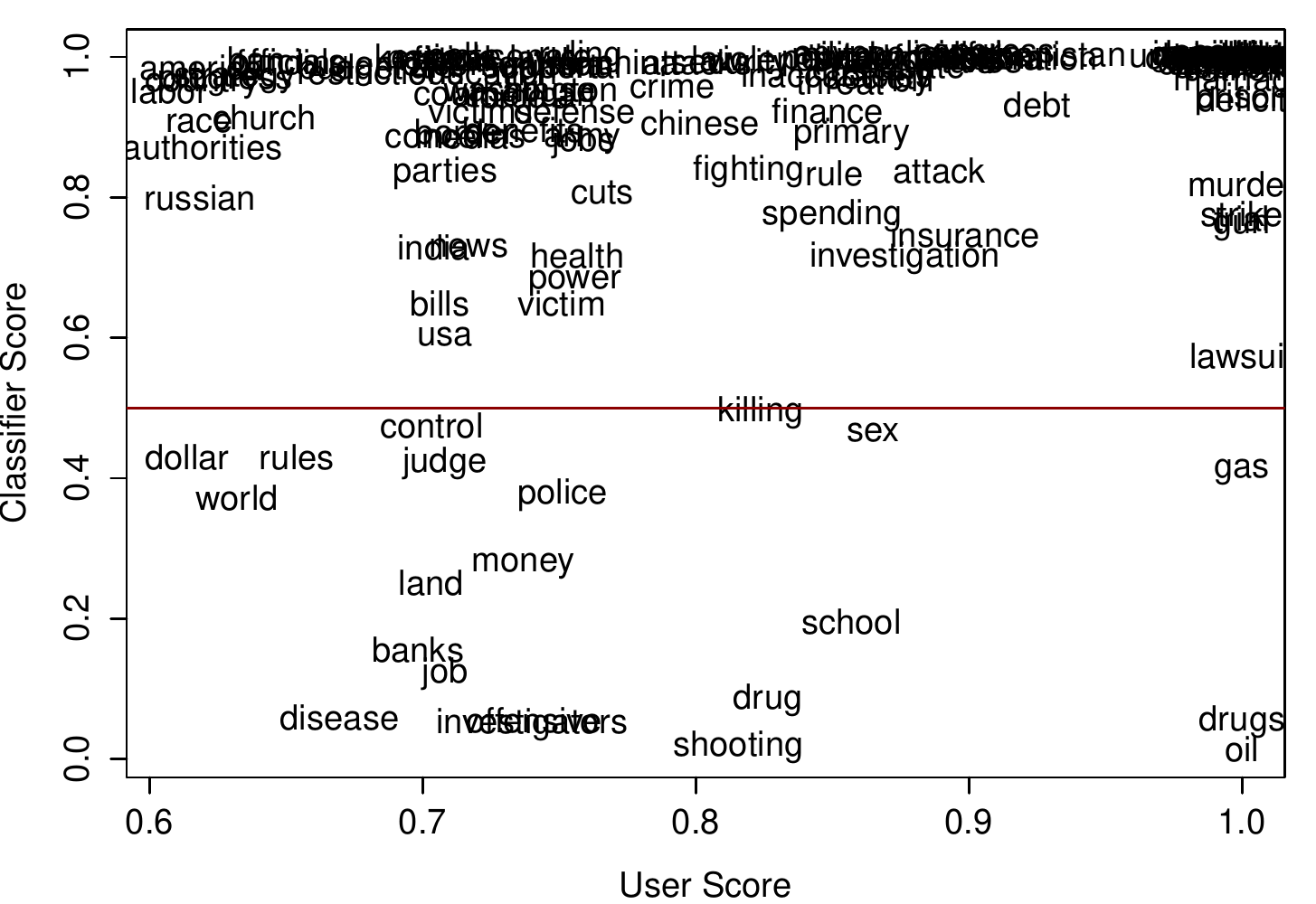}}
\caption{Scores of controversial and non-controversial words including classification errors. ``User score'' is the confidence with which the manual labeling was done (with at least 7 annotators per element), while ``classifier score'' is the output of the classifier on the training data.}
\label{fig:word-scores}
\end{figure}

Figure~\ref{fig:word-scores} depicts the training errors, which appear above the horizontal line in the plot for non-controversial topics and below it in the plot for controversial topics. To put these errors into context, we include in the figure the confidence of the manual annotation process. Note that most of the terms are classified correctly, appearing at the bottom of non-controversial and top of controversial figures. In the next section, we discuss the possible reasons for the misclassification of some of the topics.

\section{Discussion}\label{sec:discussion}

In this study, we found that controversial issues are often framed using negative emotions. Many of the terms that were found to be controversial are oftentimes related to social problems and violence, such as \term{gun}, \term{fighting}, \term{crime}, \term{victim} and \term{shooting}, or war, such as \term{strike}, \term{weapons}, and \term{army}. There may not be as many opportunities to discuss some of these issues in a non-negative light.
Other terms could be viewed in a negative, positive, or even a neutral light depending on the context at hand, for instance, terms such as \term{health}, \term{security}, or \term{jobs}. However, because these words often invoke larger ideological issues of government spending, privacy, or the economy, it may be that frames emerge that are more antagonistic. For instance, instead of emphasizing the positive aspects of one's view, a writer may choose to highlight the negative aspects of the opposing view.

Additionally, we found that controversial topics also involve less strongly emotional words. Theories of framing posit that various organizational pressures can shape the frames different news sources employ~\cite{scheufele1999framing}. This may point to an effort on the part of news agencies to more tightly control the language used when discussing controversial topics. As an overt example of organization influence on language use, many organizations today adhere to different well-known style guides when choosing certain language around sensitive topics.  The use of different style guides may point to some of the similarities and differences we saw between sources and certain controversial words. Finally, the use of biased terms become more prevalent around controversial words, with abstract notions of fairness like \term{justice} and \term{rights}, and judgment-laden terms like \term{terrorist} and \term{criminal} having a high correlation. 

Based on the above observations, we built a model to estimate to what degree a news source is treating a term as controversial. We find that some strongly controversial terms are instead classified as non-controversial---they appear in the bottom right corner of Figure~\ref{fig:word-scores-c}. In some cases, terms that have different connotations depending on the context may have lead to incorrect classifications. For instance, \term{oil}, when referenced in a financial article may treat the subject objectively as opposed to using biased or emotional language. Another example is  \term{drug} which can sometimes simply reference drug stores. However, the classification of other less ambiguous terms, such as \term{sex} and \term{killing} as non-controversial prompts re-examination of these topics as controversial. Further work is needed to model more clearly the topics represented by these terms in order to better understand their context.

The approach described in this work also allows us to examine the language of each agency around a controversy using a controlled vocabulary. For example, for the term \term{democrats} we compare the top 30 terms of the bias words lexicon used by each agency. We see that, along with more topical terms like \term{obama} and \term{democratic}, Huffington Post also includes more general and subjective terms like \term{very} and \term{good}, unlike, say, CNN and Reuters. The lexicons also allow us to glimpse a more local approach to news of smaller news agencies compared to national ones. When discussing \emph{murder}, Reuters and CNN often mention a larger framework of \term{government} and other \term{groups}, whereas Philadelphia Inquirer, Honolulu Star-Advertiser, and Houston Chronicle mention particular people (\term{woman}, \term{victim}) and places (\term{university}, \term{west}). Further development of lexical resources for deeper understanding of news coverage beyond sentiment is an exciting future direction of this research.



The large-scale analysis we have conducted is an initial inquiry into quantifying how news sources differ in their framing. By being able to pick apart how news sources differ in often subtle ways, we can begin to uncover implicit biases in framing or language use by different news entities. That many of these news sources are quite large, employing hundreds and thousands of writers and editors, suggests that organization-level pressures to conform to a particular standard or world view may in fact exist. This work can also be used to inform automatic style guide checkers, serve as a reference to journalists interested in maintaining objectivity, or readers wishing to monitor their news intake.

\spara{Data release}: our list of strongly controversial, somewhat controversial and non-controversial terms, along with their scores is available at \url{www.yelenamejova.com/resources/controversial_words.txt}. 


\spara{Acknowledgments}: the authors would like to thank NewsCred for making the data available for our research.

\balance
\bibliographystyle{nourlabbrvnat}
\bibliography{paper}



\end{document}